\documentstyle[pre,aps,aps12,mathrsfs,amsmath,tighten,graphicx]{revtex}

\begin{document}

\draft 

\title{Nonlinear resonant wave interaction in vacuum}
\author{Gert Brodin, Daniel Eriksson}
\address{Department of Plasma Physics,
Ume{\aa } University, \\ SE-901 87 Ume{\aa }, Sweden}

\author{Mattias Marklund}
\address{Department of Electromagnetics, Chalmers University of 
Technology, \\ SE--412 96 G\"{o}teborg, Sweden}

\maketitle

\ \\[-3mm] 

\begin{abstract}
The basic equations governing propagation of electromagnetic and
gravitational waves in vacuum are nonlinear. As a consequence photon-photon
interaction as well as photon-graviton interaction can take place without a
medium. However, resonant interaction between less than four waves cannot
occur in vacuum, unless the interaction takes place in a bounded region,
such as a cavity or a waveguide. Recent results concerning resonant wave
interaction in bounded vacuum regions are reviewed and extended.
\end{abstract}


\section{Introduction}

Most examples of nonlinear wave phenomena occur as a result of some
nonlinear property of a medium. In particular this applies to
electromagnetism since Maxwell's equations are linear. Still, nonlinear
interaction between photons in vacuum may occur as a result of scattering
processes involving vacuum fluctuations, as described by quantum
electrodynamics (QED). An effective theory for this can be formulated in
terms of the Euler-Heisenberg Lagrangian \cite{Heisenberg-Euler,Schwinger}.
It should be noted, however, that resonant interaction between less than
four waves requires that the waves are parallel, in which case the nonlinear
QED coupling vanishes \cite{BMS2001,Bullet}. Similarly, gravitons and
photons couple in vacuum, as described by the Einstein-Maxwell system of
equations. But the dispersion relation implies that the waves must be
parallel, in case three wave interaction should be resonant. Just as for
photon-photon interaction in vacuum, however, this condition means that the
wave-coupling vanishes \cite{Brodin-Marklund}. On the other hand, the
situation is changed if some or all of the interacting waves are confined in
bounded regions such as waveguides or cavities \cite
{Brodin-Marklund,Pegoraro1978,Reece84}. Below we will review and extend
recent results concerning resonant photon-photon scattering as well as
photon-graviton scattering in bounded regions.

\section{Photon-Photon scattering}

According to QED, the non-classical phenomenon of photon--photon scattering
can take place due to the exchange of virtual electron--positron pairs. This
give rise to vacuum polarization and magnetization currents, and an
effective field theory can be formulated in terms of the Euler--Heisenberg
Lagrangian density \cite{Heisenberg-Euler,Schwinger}
\begin{equation}
{\mathscr{L}}_{{\rm EH}}=\varepsilon _{0}{\mathscr{F}}+\xi (4{\mathscr{F}}%
^{2}+7{\mathscr{G}}^{2})\ ,  \label{eq:lagrangian}
\end{equation}
where $\xi \equiv 2\alpha ^{2}\varepsilon _{0}^{2}\hbar
^{3}/45m_{e}^{4}c^{5}$, ${\mathscr{F}}\equiv \frac{1}{2}(E^{2}-c^{2}B^{2})$,
${\mathscr{G}}\equiv c{\bf E}\cdot {\bf B}$, $e$ is the electron charge, $c$
the velocity of light, $2\pi \hbar $ the Planck constant and $m_{e}$ the
electron mass. The latter terms in (\ref{eq:lagrangian}) represent the
effects of vacuum polarization and magnetization. We note that ${\mathscr{F}}%
={\mathscr{G}}=0$ in the limit of parallel propagating waves. It is
therefore necessary to use other wave geometries in order to obtain an
effect from the QED corrections. Note that this null-result should be
expected on physical grounds, since successive Lorentz boosts along the
direction of propagation would decrease the amplitude of all waves without
limit. However, as shown in Refs. \cite{BMS2001,Bullet}, a resonant
nonlinear interaction between three waves is possible in bounded regions, as
will be considered below.

The most common approach [se e.g. Refs. \cite{valluri}-\cite{Ding}] to
calculate the interaction strength is to use the general Lagrangian, apply
the variational principle $\delta \int $ ${\mathscr{L}}_{{\rm EH}}d^{3}rdt=0$
to find the influence of the QED terms in Maxwells equations (where the
Lagrangian should be expressed in terms of the 4-potential), and proceed
from there using standard techniques for weakly nonlinear waves. However, as
shown in Ref. \cite{Bullet}, the algebra is significantly reduced if one
make an ansatz for the potentials corresponding to the field geometry of
consideration, and then derive the evolution equations directly from the
variational principle, without using Maxwells equations as an intermediate
step.

Following the later approach we start by considering three interacting waves
in a cavity with the shape of a rectangular prism. We let one of the corners
lie in the origin, and we let the opposite corner have coordinates $%
(x_{0},y_{0},z_{0})$. In practice we are interested in a shape where $%
z_{0}\gg x_{0},y_{0}$ but this assumption will not be used in the
calculations. We let the large amplitude pump modes have vector potentials
of the form
\begin{equation}
{\bf A}_{1}=A_{1}\sin \left( \frac{\pi x}{x_{0}}\right) \sin \left( \frac{%
n_{1}\pi z}{z_{0}}\right) \exp (-{\rm i}\omega _{1}t)\widehat{{\bf y}}+{\rm %
c.c}  \label{vector-rect1}
\end{equation}
and
\begin{equation}
{\bf A}_{2}=A_{2}\sin \left( \frac{\pi y}{y_{0}}\right) \sin \left( \frac{%
n_{2}\pi z}{z_{0}}\right) \exp (-{\rm i}\omega _{2}t)\widehat{{\bf x}}+{\rm %
c.c.}  \label{vector-rect2}
\end{equation}
where ${\rm c.c.\,}$\ denotes complex conjugate, and we chose the radiation
gauge $\ \Phi =0$. It is easily checked that the corresponding fields
\begin{mathletters}
\label{eq:pump10}
\begin{eqnarray}
B_{1z} &=&\left( \frac{\pi }{x_{0}}\right) A_{1}\,\cos \left( \frac{\pi x}{%
x_{0}}\right) \sin \left( \frac{n_{1}\pi z}{z_{0}}\right) \exp (-{\rm i}%
\omega _{1}t)+{\rm c.c.}\ ,  \label{FPR11} \\
B_{1x} &=&-\left( \frac{n_{1}\pi }{z_{0}}\right) A_{1}\sin \left( \frac{\pi x%
}{x_{0}}\right) \cos \left( \frac{n_{1}\pi z}{z_{0}}\right) \exp (-{\rm i}%
\omega _{1}t)+{\rm c.c.}\ ,  \label{FPR12} \\
E_{1y} &=&{\rm i}\omega _{1}A_{1}\sin \left( \frac{\pi x}{x_{0}}\right) \sin
\left( \frac{n_{1}\pi z}{z_{0}}\right) \exp (-{\rm i}\omega _{1}t)+{\rm c.c.}%
\ ,  \label{FPR13}
\end{eqnarray}
together with $\omega _{1}^{2}=n_{1}^{2}\pi ^{2}c^{2}/z_{0}^{2}+\pi
^{2}c^{2}/x_{0}^{2}$, and
\end{mathletters}
\begin{mathletters}
\label{eq:pumpny10}
\begin{eqnarray}
B_{2z} &=&-\left( \frac{\pi }{y_{0}}\right) A_{2}\,\cos \left( \frac{\pi y}{%
y_{0}}\right) \sin \left( \frac{n_{2}\pi z}{z_{0}}\right) \exp (-{\rm i}%
\omega _{2}t)+{\rm c.c.}\ ,  \label{FPR21} \\
B_{2y} &=&\left( \frac{n_{2}\pi }{z_{0}}\right) A_{2}\sin \left( \frac{\pi y%
}{y_{0}}\right) \cos \left( \frac{n_{2}\pi z}{z_{0}}\right) \exp (-{\rm i}%
\omega _{2}t)+{\rm c.c.}\ ,  \label{FPR22} \\
E_{1x} &=&{\rm i}\omega _{2}A_{2}\sin \left( \frac{\pi y}{y_{0}}\right) \sin
\left( \frac{n_{2}\pi z}{z_{0}}\right) \exp (-{\rm i}\omega _{2}t)+{\rm c.c.}%
\ ,  \label{FPR23}
\end{eqnarray}
together with $\omega _{2}^{2}=n_{2}^{2}\pi ^{2}c^{2}/z_{0}^{2}+\pi
^{2}c^{2}/y_{0}^{2}$ \ are proper eigenmodes fulfilling Maxwells equations
and the standard boundary conditions. Similarly we assume that the mode to
be excited can be described by a vector potential
\end{mathletters}
\begin{equation}
{\bf A}_{3}=A_{3}\sin \left( \frac{\pi y}{y_{0}}\right) \sin \left( \frac{%
n_{3}\pi z}{z_{0}}\right) \exp (-{\rm i}\omega _{3}t)\widehat{{\bf x}}+{\rm %
c.c.}  \label{vector-rect3}
\end{equation}
where $\omega _{3}^{2}=n_{3}^{2}\pi ^{2}c^{2}/z_{0}^{2}+\pi
^{2}c^{2}/y_{0}^{2}$, in which case we get fields of the same form as in
Eqs. (\ref{FPR21})-(\ref{FPR23}). Since the QED terms are fourth order in
the amplitude, the corresponding nonlinearities are cubic, implying that $%
\omega _{3}=\pm (2\omega _{2}\pm \omega _{1})$ or \ $\omega _{3}=\pm (\omega
_{2}\pm 2\omega _{1})$ should hold for resonant interaction. We assume that
the cavity dimensions are chosen such that a single eigenmode can be
resonantly excited, and pick the alternative
\begin{equation}
\omega _{3}=2\omega _{1}-\omega _{2}  \label{Frequency matching}
\end{equation}
for definiteness. We note that when performing the variations $\delta
A_{3}^{\ast }$, the lowest order terms proportional to $\delta A_{3}^{\ast
}A_{3}$ vanish due to the dispersion relation, and we need to include terms
due to the time dependence of the amplitude of the type $A_{3}\partial
(\delta A_{3}^{\ast })/\partial t$. For the fourth order QED corrections
proportional to $\delta A_{3}^{\ast }$, only terms proportional to $%
A_{1}^{2}A_{2}^{\ast }\delta A_{3}^{\ast }$ survives the time integration,
due to the frequency matching (\ref{Frequency matching}). After some algebra
the corresponding evolution equation for mode 3 becomes:
\begin{equation}
\left( \frac{d}{dt}-\gamma \right) A_{3}=\frac{\varepsilon _{0}\kappa \omega
_{3}^{3}}{8{\rm i}}K_{{\rm rec}}A_{1}^{2}A_{2}^{\ast }  \label{Evolution1}
\end{equation}
where the dimensionless coupling coefficient $K_{{\rm rec}}$ is
\begin{eqnarray}
K_{{\rm rec}} &=&\frac{1}{\omega _{3}^{4}}\left\{ -(-,+)2\omega
_{1}^{2}\omega _{2}\omega _{3}+2\pi c^{4}\left[ -(-,+)\frac{3}{%
x_{0}^{2}y_{0}^{2}}+\frac{n_{1}^{2}n_{2}n_{3}}{z_{0}^{4}}+(+,-)\frac{%
n_{1}^{2}}{y_{0}^{2}z_{0}^{2}}\right. \right.  \nonumber \\
&&\left. -\frac{n_{2}n_{3}}{x_{0}^{2}z_{0}^{2}}\right] +\pi ^{2}c^{2}\left[
-(-,+)\frac{5n_{1}^{2}\omega _{2}\omega _{3}}{z_{0}^{2}}+\frac{5\omega
_{1}^{2}n_{2}n_{3}}{z_{0}^{2}}\right.  \nonumber \\
&&\left. \left. -(-,+)\left( \frac{2\omega _{2}\omega _{3}}{x_{0}^{2}}+\frac{%
2\omega _{1}^{2}}{y_{0}^{2}}\right) +-(+,-)\frac{7\omega _{1}\omega
_{2}n_{2}n_{3}}{z_{0}^{2}}+(-,-)\frac{7\omega _{1}\omega _{3}n_{1}n_{2}}{%
z_{0}^{2}}\right] \right\}  \label{Coupling1}
\end{eqnarray}
and we have added a phenomenological linear damping term represented by $%
\gamma $. When deriving Eq. (\ref{Coupling1}) we have assumed one of the
following mode number matchings
\begin{eqnarray}
2n_{1}-n_{2}+n_{3} &=&0  \label{mode1} \\
2n_{1}+n_{2}-n_{3} &=&0  \label{mode2} \\
2n_{1}-n_{2}-n_{3} &=&0  \label{mode3}
\end{eqnarray}
in order for the QED corrections terms to survive the z-integration. The
three different sign alternatives in (\ref{Coupling1}) correspond to the
mode number matching options (\ref{mode1})-(\ref{mode3}) respectively. Given
experimental data for possible values of the pump field strengths and the
damping coefficient inside cavities, the saturation level of the excited
mode can be determined from Eq. (\ref{Evolution1}). The possibilities for
detection of photon-photon scattering using currently available performance
on microwave cavities will be discussed in the final section of the paper.

\section{Gravitational interaction}

{\em Preliminaries.} In vacuum, a linearized gravitational wave can be
transformed into the transverse and traceless (TT) gauge. Then we have the
following line-element
\begin{eqnarray}
{\rm d}s^{2} &=& -c^2{\rm d}t^2 + {\rm d}x^2 + [1 + h_+(\xi)]\,{\rm d}y^2
\nonumber \\
& &\quad + [1 - h_+(\xi)]\,{\rm d}z^2 + 2h_{\times}(\xi)\,{\rm d}y\,{\rm d}z
\ ,  \label{le}
\end{eqnarray}
where $\xi \equiv x-ct$, $c$ is the speed of light in vacuum, and $|h_{+}|,
|h_{\times}| \ll 1$.

Neglecting terms proportional to derivatives of $h_{+}$ and $h_{\times }$
(since the gravitational frequency is assumed small), the wave equation for
the magnetic field is \cite{Brodin-Marklund}
\begin{equation}
\left[ \frac{1}{c^{2}}\frac{\partial ^{2}}{\partial t^{2}}-\nabla ^{2}\right]
{\boldsymbol{B}}=\left[ h_{+}\left( \frac{\partial ^{2}}{\partial y^{2}}-%
\frac{\partial ^{2}}{\partial z^{2}}\right) +h_{\times }\frac{\partial ^{2}}{%
\partial y\partial z}\right] {\boldsymbol{B}},  \label{WaveB}
\end{equation}
and similarly for the electric field.

In an isotropic dielectric medium with permittivity $\varepsilon $ different
from the vacuum permeability, the equation (\ref{WaveB}) still holds, simply
if we replace $c$ by $c/n=c/\sqrt{\varepsilon _{r}}$ in the above
expressions, where $\varepsilon =\varepsilon _{r}\varepsilon _{0}$. For the
moment, we will neglect mechanical effects, i.e., effects which are
associated with the varying coordinate distance of the cavity walls due to
the gravitational wave.

{\em Cavity design.} The coupling of two electromagnetic modes and a
gravitational wave in a cavity will depend strongly on the geometry of the
electromagnetic eigenfunctions. It turns out that we can greatly magnify the
coupling, as compared to a rectangular prism geometry, by varying the
cross-section of the cavity, or by filling the cavity partially with a
dielectric medium. The former case is of more interest from a practical
point of view, since a vacuum cavity implies better detector performance,
but we will here consider the later case since it can be handled
analytically.

Specifically, we choose a rectangular cross-section (side lengths $x_0$ and $%
y_0$), and we divide the length of the cavity into three regions. Region 1
has length $l_1$ (occupying the region $-l_1<z<0$) and a refractive index $%
n_1$. Region 2 has length $l_2$ (occupying the region $0<z<l_2$), with a
refractive index $n_2$, while region 3 consists of vacuum and has length $%
l_3 $ (occupying the region $l_2 < z < l_3 + l_2$). We will also use $l =
l_1 + l_2 + l_3$ for the total length. The cavity is supposed to have
positive coordinates, with one of the corners coinciding with the origin.
Furthermore, we require that $l_2 \ll l_1$, and that the wave number in
region 2 is less than in region 1. The reason for this arrangement is
twofold. Firstly, we want to obtain a large coupling between the wave modes,
and secondly we want an efficient filtering of the eigenmode with the lower
frequency in region three.

The first step is to analyze the linear eigenmodes in this system. Those
with the lowest frequencies are modes of the type
\begin{mathletters}
\label{Region1}
\begin{eqnarray}
E_{y} &=&\frac{i\omega x_{0}}{m\pi }\widetilde{B}_{zj}\sin \left( \frac{m\pi
x}{x_{0}}\right) \sin [k_{j}z+\varphi _{j}]e^{-i\omega t}, \\
B_{z} &=&\widetilde{B}_{zj}\cos \left( \frac{m\pi x}{x_{0}}\right) \sin
[k_{j}z+\varphi _{j}]e^{-i\omega t}, \\
B_{x} &=&-\frac{k_{j}x_{0}}{m\pi }\widetilde{B}_{zj}\sin \left( \frac{m\pi x%
}{x_{0}}\right) \cos [k_{j}z+\varphi _{j}]e^{-i\omega t},
\end{eqnarray}
in regions $j=1$, $2$ and $3$, respectively, where the wave in region 3 is a
standing wave. Furthermore, in region 3 we may also have a decaying wave
\end{mathletters}
\begin{mathletters}
\label{Region2b}
\begin{eqnarray}
E_{y} &=&\frac{i\omega x_{0}}{m\pi }\widetilde{B}_{z3}\sin \left( \frac{m\pi
x}{x_{0}}\right) \sinh [k_{3}z+\varphi _{3}]e^{-i\omega t}, \\
B_{z} &=&\widetilde{B}_{z3}\cos \left( \frac{m\pi x}{x_{0}}\right) \sinh
[k_{3}z+\varphi _{3}]e^{-i\omega t}, \\
B_{x} &=&-\frac{k_{3}x_{0}}{m\pi }\widetilde{B}_{z3}\sin \left( \frac{m\pi x%
}{x_{0}}\right) \cosh [k_{3}z+\varphi _{3}]e^{-i\omega t}.
\end{eqnarray}
Using standard boundary conditions, it is straightforward to perform most of
the eigenmode calculations analytically. Once the wavenumbers are calculated
for an eigenmode, the relation between the amplitudes in the three regions
is found, and thereby the mode profile. We are specifically interested in
the shift from decaying to oscillatory behavior in region 3, and we denote
the highest frequency which is decaying in region 3 with index $a$, and the
wave number and decay coefficient with $k_{1a}$, $k_{2a}$ and $k_{3a}$
respectively. Similarly, the next frequency, which is oscillatory in both
regions, is denoted by index $b$. If we have $l\gg x_{0}$ (and $m$ the same)
these two frequencies will be very close, and a gravitational wave which has
a frequency equal to the difference between the electromagnetic modes causes
a small coupling between these modes. An example of two such eigenmodes is
shown in fig.\ 1.

\begin{figure}[ht]
\begin{center}
  \includegraphics[width=.9\textwidth]{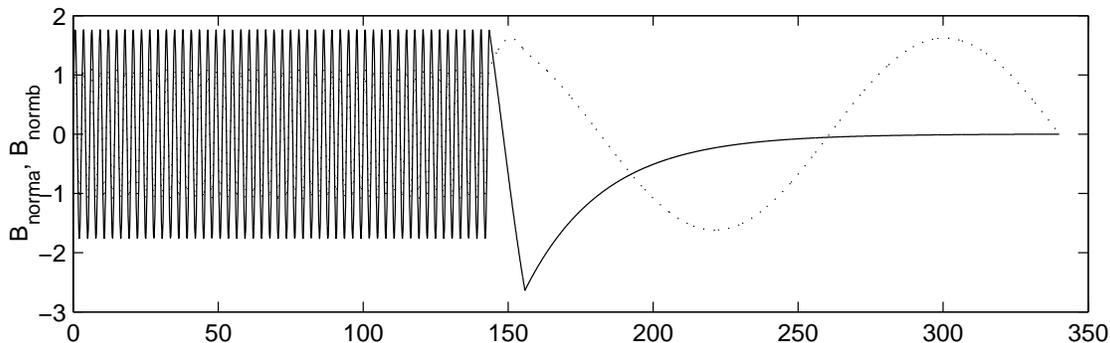}
\end{center}
\caption{Mode profiles for the eigenmodes are shown for the parameter
values $n_{1}$ $=1.22$, $n_{2}=1.0005$, $l_{1}m/x0=143$, $l_{2}m/x0=12.3$
and $l_{3}m/x0=184$. $B_{{\rm norm}a}$ is the solid line and $B_{{\rm norm}%
b} $ is the dotted line, and $I_{c}=0.27$.}
\end{figure}

We define the eigenmodes to have the form $\widetilde{B}_{za,b}=\widetilde{B}%
_{a,b}(t)B_{{\rm norm}a,b}(z,y,z)$, where $\widetilde{B}_{a,b}$ is a
time-dependent amplitude and the normalized eigenmodes $B_{{\rm norm}a,b}$
fulfill $\int_{V}\left| B_{{\rm norm}a,b}\right| ^{2}dV=V$. We let all
electromagnetic field components be of the form $A=A_{a}(\boldsymbol{r})\exp
(-i\omega _{a}t)+A_{b}(\boldsymbol{r})\exp (-i\omega _{b}t)+{\rm c.c.}$,
where c.c.\ stands for complex conjugate, and the indices stand for the
eigenmodes discussed above. The gravitational perturbation can be
approximated by $h_{+,\times }=\widehat{h}_{+,\times }\exp (-i\omega _{g}t)+%
{\rm c.c}$, where we neglect the spatial dependency of $\widehat{h}%
_{+,\times }$ altogether, since the gravitational wavelength is assumed to
be much longer than all of the cavity dimensions. If we consider a binary
system of two black holes fairly close to collapse, the gravitational
frequency will not be an exact constant, but will increase slowly. During a
certain interval in time, the frequency matching condition
\end{mathletters}
\begin{equation}
\omega _{b}=\omega _{a}+\omega _{g},  \label{Freq-match}
\end{equation}
will be approximately fulfilled. Given the wave equation (\ref{WaveB}), and
the above ansatz we find after integrating over the length of the cavity
\begin{equation}
\frac{2i\omega _{b}}{c^{2}}\left( \frac{\partial }{\partial t}-\gamma
\right) \widetilde{B}_{b}=-h_{+}k_{1a}^{2}\widetilde{B}_{a}I_{c},
\label{Excitation-eq}
\end{equation}
where
\begin{equation}
I_{c}=\frac{1}{Vk_{1a}^{2}}\int_{V}\frac{\partial ^{2}B_{{\rm norm}\,a}}{%
\partial z^{2}}B_{{\rm norm}\,b}\,dV,  \label{Coupling-int}
\end{equation}
and we have added a phenomenological linear damping term represented by $%
\gamma $. Thus we note that for the given geometry, only the $h_{+}$%
-polarization gives a mode-coupling. (Rotating the cavity $\pi /4$ around
the $x$-axis will instead give coupling to the $h_{\times }$-polarization.)
Furthermore, if we consider propagation in a different angle to the cavity ,
the result will be slightly modified. Calculations of the eigen-mode
parameters show that $I_{c}$ may be different from zero when $n_{1,2}\neq 1$%
, and generally $I_{c}$ of the order of unity can be obtained, see fig.\ 1
for an example. From Eq.\ (\ref{Excitation-eq}), we find that the saturated
value of the gravitationally excited mode is
\begin{equation}
\left| \widetilde{B}_{b{\rm sat}}\right| =\frac{h_{+}k_{1a}^{2}c^{2}%
\widetilde{B}_{a}}{2\gamma \omega _{b}}I_{c}\ .  \label{Saturation-eq}
\end{equation}
In fig.\ 1 it is shown that we can get an appreciable mode-coupling constant
$I_{c}$ for a cavity filled with materials with different dielectric
constants, and it is of much interest whether the same can be achieved in a
pure vacuum cavity. As seen by Eq.\ (\ref{Coupling-int}), the coupling is
essentially determined by the wave numbers of the modes, given by $%
k=(m^{2}\pi ^{2}/x_{0}^{2}-n^{2}\omega ^{2}/c^{2})^{1/2}$. Thus by adjusting
the width $x_{0}$ in a vacuum cavity, we may get the same variations in the
wave numbers as when varying the index of refraction $n$. The translation of
our results to a vacuum cavity with a varying width is not completely
accurate, however. Firstly, when varying $x_{0}$, the mode-dependence on $x$
and $z$ does not exactly factorize, in particular close to the change in
width. Secondly, the contribution to the coupling $I_{c}$ in each section
becomes proportional to the corresponding volume, and thereby also to the
cross-section. However, since most of the contribution to the integral in
Eq.\ (\ref{Coupling-int}) comes from region 1, our results can still be
approximately translated to the case of a vacuum cavity, by varying $x_{0}$
instead of $n$ such as to get the same wavenumber as in our above example.
Thus we conclude that our discussion of the sensitivity of \ a cavity based
detector given below, can be based on a vacuum cavity rather than a
''dielectric cavity'', where the former case is preferred due to the much
smaller dissipation level.

\section{Discussion of the detection sensitivity}

For both cases considered above, as can be seen from equations (\ref
{Evolution1}) and (\ref{Saturation-eq}) respectively, the quality factor $Q$
of the cavity (i.e. the damping time/periodtime) and the allowed field
strength are the main parameters that determine the saturation level of the
excited mode. For superconducting niobium cavities large surface field
strengths ($E\sim 60$ ${\rm MV/m}$) and high quality factors $Q\sim 10^{10%
\text{ }}$can be obtained simultaneously \cite{Performance}. For the case of
photon-photon scattering this typically corresponds to a few excited
microwave photons in the new mode. Provided the number of excited photons
beats the thermal noise level ($N_{th}\sim kT/\hbar \omega _{3}$) a few
microwave photons is enough for detection, see e.g. \cite{Walther2001} .
However, detection of such a small signal can only take place if the pump
waves are filtered out. This can be done directly in the cavity using a
''filtering geometry'', i.e. a cavity with a variable cross-section, where
only the excited mode has a frequency above cut-off in the detection region
of the cavity. Our conclusion is that detection of photon-photon scattering
is possible in a microwave cavity using current technology, provided the
equipment has state of the art performance, and a filtering geometry is
applied.

The considerations for the gravitational wave detector are similar to the
photon-photon scattering case. However, we must also take the length
variations of the detector due to acoustic thermal noise into account. As a
consequence, an efficient detector must then have a rather high mechanical
quality factor $Q_{{\rm mec}}\approx 6\times 10^{6}$(where $Q_{{\rm mec}}$
describes the relative damping of acoustic oscillations), for sensitive
detection to be possible. Furthermore, we must also be aware that the
gravitational wave sources of most interest - binary systems close to
collapse - produces radiation with a finite frequency chirp, implying a
finite time of coherent interaction, typically fractions of a second. If the
above aspects are taken into account, we find that a cavity based detector
of a few meters length can detect metric perturbations of the order $h_{\min
}\approx $ $7\times 10^{-23}$(See Ref. \cite{Brodin-Marklund} for a more
thorough discussion of the detection level.).

\end{document}